\documentclass{iopart}

\usepackage{epsfig}

\newcommand{\be}{\begin{equation}}
\newcommand{\ee}{\end{equation}}
\newcommand{\bea}{\begin{eqnarray}}
\newcommand{\eea}{\end{eqnarray}}
\newcommand{\non}{\nonumber}

\begin{document}

\title{Observational signatures of pre-inflationary and lower-dimensional effective gravity}

\author{Massimiliano Rinaldi}

\address{D\'epartment de Physique Th\'eorique, Universit\'e de Gen\`eve, \\ 24 quai E.\ Ansermet  CH--1211 Gen\`eve 4, Switzerland.}  
\ead{massimiliano.rinaldi@unige.ch}

\begin{abstract}
Several alternative ways to quantize gravitational interactions seem to indicate that gravity at short distances is effectively two-dimensional. If this were true, and the energy scale at which the dimensional transition occurs is of the order of the inflationary one, it is reasonable to investigate eventual signatures in the primordial perturbation spectra. In this paper we look at this possibility by assuming that the inflationary era was preceded by an effectively two-dimensional evolution of the Universe. We model the transition with a mode matching and we show that, in this case, no observational signatures are expected in the tensor and scalar power spectra nor in their ratio. We also show that assuming modified dispersion relations before the matching between two sets of four-dimensional modes leads to similar results. 
\end{abstract}

\section{Introduction}

The inflationary theory predicts that tiny quantum fluctuations of the metric and of the inflaton field have evolved into the seeds of the large scale structures that we see today in the sky \cite{slava,staro}. This striking connection allows in principle to investigate the physical properties of our Universe at the beginning of its own existence, when quantum gravity effects are expected to be important. As several coming experiments, such as Planck, are expected to provide for very precise data on the cosmic microwave background (CMB), we are facing the possibility of exploring the quantum gravitational realm for the very first time. However, as a fundamental quantum theory of gravitation is lacking, we do not really know what to expect from the new data. It is therefore very important to elaborate various predictions to be confronted with the  measurements. 

The question whether the effects arising in the planckian era are too small to be observed in the CMB or via gravitational waves is still open, and several authors have expressed opposite views, see e.\ g.\ \cite{transP} . These conclusions were mainly drawn by considering short scale modifications of gravity, where Lorentz-breaking mechanism were at work via modified dispersion relations. There exist however other possibilities. For example,   the standard renormalization of quantum fields on curved space might lead to observable effects without resorting to trans-planckian physics \cite{renorm}.

On more fundamental grounds, several attempts to quantize gravity consistently lead to an intriguing common indication, namely that at short distance, gravity is effectively two-dimensional (see e.\ g.\ \cite{carlip} for a review and \cite{others} for further proposals). If we take this hint seriously, the next step is to look at possible experimental signatures, in particular in cosmological observations. In fact, if the CMB inhomogeneities have originated from stretched short-distance quantum fluctuations occurred during the inflationary era, it is well possible that they have preserved some memory of a fundamental two-dimensional structure of space-time. 

In this paper we wish to look closely at this idea but in a model-independent way. In other words, we do not choose any particular model of quantum gravity, but we simply assume that scalar and tensor perturbations observed today originated from an effectively two-dimensional perturbation at a certain time $t_{i}$, which corresponds to $N$ e-folds before the horizon exit of the scales of interest. Our aim is to see if current observations are affected by such a scenario. In this case, we would like to determine if data would be able to put some constraint on the value of $N$. In order to do so, we proceed by matching modes emerging from a two-dimensional Bunch and Davies vacuum to four-dimensional modes evolving towards the horizon exit through a slowly rolling inflationary Universe. Certainly, this is a crude approximation of a process, which probably is much more complicated and model-dependent. Furthermore, one might be concerned by the possibility that a dimensional transition from two to four dimensions creates a preferred direction in space-time. However, as the Universe inflates quasi exponentially after the match, we know that anisotropies will rapidly decay, and the space-time becomes locally de-Sitter \cite{wald}, well before the horizon exit of cosmological scales of interest.  Alternatively, we could also take on-board the suggestion, made in \cite{carlip2}, that each point of space-time picks a random preferred direction. In the vicinity of the point, the effective metric is a Kasner space with random coefficients, that rapidly decay into a locally flat metric. 

Mode matching is a technique often invoked in order to eliminate the infrared (IR) divergences that occurs in the power spectrum, se e.\ g.\ \cite{proko}. According to this procedure, one assumes that the scale factor describes a radiation-dominated (or a Minkowski space-time) Universe for $t<t_{i}$, and a quasi exponentially expanding Universe for $t>t_{i}$. The only requirement is that both the scale factor and its time-derivative are continuous at $t_{i}$. As a result, the two-point function (at coincident points) of a test massless scalar field is IR-finite at all times. The same conclusion can be extended to the power spectrum of tensor and scalar perturbations. Such a construction brings in the new parameter $t_{i}$ that, in principle, can be constrained by the observed values of, for example, the spectral indices. However, as we will show below, the impact of the matching on the spectral indices vanishes, and the pre-inflationary phase is basically unobservable. Mode matching was also used to investigate gravitational waves eventually generated by sudden matter-radiation transition, and the results showed no detectable modifications of the power spectrum \cite{maxruth}.

This conclusion does not necessarily hold if the initial conditions for the four-dimensional perturbations are set by the fluctuations of a scalar field defined on a two-dimensional space-time. As we will show below, the different dimensionality of the wave functions implies that the spectral indices depend on the time at which the matching takes place. We also anticipate that a similar result holds in the case when, for $t<t_{i}$, one  allows for superluminal high-frequency dispersion relations without changing the number of spatial dimensions.

The plan of the paper is the following. In the next section we present the formalism that we use to implement the matching conditions between modes. Then, we apply it to determine the matching conditions of the tensor and the scalar perturbations, and to calculate the corresponding spectra in the case when dimensionality is preserved in Sec.\ \ref{modesmatch}. In Sec.\ \ref{Dtrans} we turn to the case of a transition between two and four dimensions, and we compute  the tensor-to-scalar ratio and the spectral indices. In this section we  present an explicit model of transition, that allows for scalar and tensor perturbations to be well-defined also before the matching.  We discuss these results in Sec.\ \ref{conc} and conclude with some considerations and prospects for future investigations.


\section{Tensor and scalar perturbations}\label{TSpert}


We begin our investigation by briefly recalling the properties of the mode equation for a test massless scalar field on a $D$-dimensional Friedmann-Lema\^{\i}tre-Robertson-Walker (FLRW) background, according to the conventions of \cite{proko}. In the following we set $c=\hbar=1$ and we denote by $M_{\rm pl}$ the Planck mass and with $G=(8\pi M_{\rm pl}^{2})^{-1}$ the Newton constant. Let us consider the metric  
\bea\label{metric}
 ds^2=-dt^2+a^2(t)\delta_{ij}dx^idx^j\ ,
 \eea
 where $i,j$ run from 1 to $(D-1)$. A massless scalar field $\phi$ on this background satisfies the Klein-Gordon equation 
  \bea\label{KG}
 (\nabla_{\mu}\nabla^{\mu}-\xi R)\phi(x)=0\ .
 \eea
where $\xi$ is the coupling of the field to the Ricci scalar, which can be expressed as
 \bea
 R=(D-1)(D-2\epsilon)H^2 \ , \quad \epsilon=-{\dot H\over H^2}\ .
 \eea
 The Fourier transform of $\phi$, defined by
 \bea
 \tilde \phi(k)=\int d^{D-1}x\,e^{-i\vec k\cdot \vec x}\phi(x)\ ,
 \eea
 obeys the equation
 \bea\label{fmodeeq}
 \ddot{\tilde\phi}+(D-1)H\dot{\tilde\phi}+\left({k^2\over a^2}+\xi R\right)\tilde\phi=0 \ ,
 \eea
 where the dot stands for differentiation with respect to $t$.
 Upon quantization, we promote $\tilde\phi$ to the operator
 \bea
 \hat \phi=\psi(t,k)\hat b+\psi^*(t,k)\hat b^{\dagger}\ ,\quad k\equiv |\vec{k}|\ ,
 \eea
 such that
 \bea
 [\hat b(\vec k),\hat b^{\dagger}(\vec k')]=(2\pi)^{D-1}\delta(\vec k-\vec k ')\ .
 \eea
 In turn, this implies the canonical commutation rule
 \bea
 [\hat \phi(t,\vec k),a^{D-1}\dot{\hat \phi}^{\dagger}(t,\vec k')]=i(2\pi)^{D-1}\delta(\vec k-\vec k ')\ ,
 \eea
 and the Wronskian condition
 \bea\label{wronsk}
 \psi\dot\psi^*-\psi^*\dot\psi=ia^{1-D}\ .
 \eea
In the following, we consider only space-times with constant $\epsilon$ as they are sufficient to describe both inflationary solutions and matter-dominated universes.  In this case, the modes satisfy the equation
 \bea\label{modeseq}
 \left(z^2{d^2\over dz^2}+z^2+{1\over 4}-\mu^2  \right)\left(a^{{D\over 2}-1}\psi\right)=0\ ,
 \eea
 where we have defined the variable
 \bea\label{zvar}
 z(t,k)={k\over (1-\epsilon)\dot a(t)}\ ,
 \eea
 and where
 \bea
 \mu^2=\left[D-1-\epsilon\over 2(1-\epsilon)\right]^2-\xi\,{(D-1)(D-2\epsilon)\over (1-\epsilon)^2}\ .
 \eea
With these settings, a radiation-dominated Universe corresponds to $\epsilon=D/2$ so $\mu=1/2$.
The general solution to Eq.\ (\ref{modeseq}) can be written as
  \bea\label{psisoln}
 \psi(t,k)=a^{1-{D\over 2}}(t)\left[E_ku(z)+F_ku^*(z)\right]\ ,
 \eea
 with
 \bea\label{umode}
 u(z)=\sqrt{\pi z\over4k}H^{(1)}_\mu(z)\ ,
 \eea
 where $H^{(1)}_{\mu}$ is the Hankel function of first kind.  The wronskian condition (\ref{wronsk}) implies that $|E_k|^2-|F_k|^2=1$. To recover the Minkowskian solution, one has to take $z$ large, so that the large argument approximation of the Hankel function \cite{abramo} yields
\bea\label{largearg}
u_{\mu}(z_{i})\simeq {1\over \sqrt{2k}}\exp \left[i\left(z_{i}-{\mu\pi\over 2}-{\pi\over 4}\right)\right]\ .
\eea
 
The detailed analysis of the evolution of a massless scalar field is particularly important, as metric perturbations, in four or more dimensions, are known to satisfy the same equation as (\ref{modeseq}). To see this it is sufficient to consider the perturbed metric
 \bea
 ds^{2}=-dt^{2}+a^{2}(t)(\delta_{ij}+h_{ij})dx^{i}dx^{j}\ ,
 \eea
 and expand the field $h_{ij}$ in plane-wave modes $h(t)e_{ij}e^{i\vec k\cdot \vec x}$, where $e_{ij}$ is the constant polarization vector. Then, one can easily show that $h(t)$ satisfies (\ref{fmodeeq}), and the associated tensor power spectrum is defined as
 \bea
 P_{T}=16\pi k^{3}|h(t)|^{2}\ .
 \eea
In standard (single-field) slow-roll inflation, $\epsilon$ is a small parameter related to the inflaton potential $V(\Phi)$ via $\epsilon=(M_{\rm pl}^{2}/2)(V'/V)^{2}$. In this case, the power spectrum is nearly scale-invariant as $P_{T}\propto k^{-2\epsilon}$.
 
The scalar perturbations are defined as a combination of metric and inflaton fluctuations. Normally, one separates the inflaton field into a homogeneous part plus a tiny fluctuation, according to $\Phi(t,\vec x)=\Phi_{0}(t)+\delta\Phi(t,\vec x)$, where $\delta\Phi(t,\vec x)$ is decomposed into plane waves as for $h_{ij}$. Then, the Fourier transform  of the perturbation, in terms of the Mukhanov variable  $\zeta(t,k)$,  is described  by the equation \cite{muk}
\bea
\ddot {\zeta}+(D-1)H\dot {\zeta}+\left[3\eta H^2-2(D-1)\epsilon H^2+{k^2\over a^2}\right]\zeta=0\ ,
\eea
where $\eta=M_{\rm pl}^2(d^{2}V/d\Phi^{2})/V$  is the second order slow-roll parameter and $V$ the inflaton potential. This equation holds in the slow-roll approximation, and can be mapped into Eq.\ (\ref{fmodeeq}) by means of the identification
\bea\label{iden}
\xi = {3\eta-2(D-1)\epsilon\over (D-1)(D-2\epsilon)}\ ,
\eea
thus the general solution is again given by the expressions (\ref{psisoln}) and (\ref{umode}), but with the index given by 
\bea
\mu={(D-1)\over 2}+{(D^{2}+D-2) \epsilon\over 2(D-1)}-{3\eta\over (D-1)}\ .
\eea
Finally, in four dimensions the scalar power spectrum is defined as
\bea
P_{S}=4\pi k^{3}\left(H\over \dot\Phi_{0}\right)^{2}|\zeta|^{2}\ ,
\eea
where the slow-roll condition allows to estimate $(\dot\Phi_{0}/H)\simeq 2M_{\rm pl}^{2}\epsilon $. Consistency also requires that $\eta\ll 1$. In single-field inflation, one finds a nearly scale-invariant spectrum of the form $P_{S}\propto k^{n_{s}-1}$ with $n_{s}-1=2\eta-6\epsilon\,\,$.

Some care needs to be taken in the case $D=2$. The solution (\ref{psisoln}) to equation (\ref{modeseq}) is still valid, and reduces to plane waves when $\xi=0$, as the Hankel function is proportional to $z^{-1/2}\exp(iz)$ when $\mu=1/2$. This is a consequence of the conformal invariance of the wave equation when minimally coupled to gravity. In two dimensions, minimal coupling corresponds precisely to $\xi=0$ \cite{BirDav}. Of course, when $D=2$, there are no tensor modes, and Eq.\ (\ref{modeseq}) describes the propagation of a scalar degree of freedom only. As we will see in Section 4, things are different when a time-dependent space-time is four-dimensional all the time, but two spatial directions are compact and non-dynamical up to a critical time $t_0$, after which they decompactify and begin to evolve. In this case, tensor perturbations are present and well-defined also for $t<t_0$.

 
 \section{Mode-matching as an infrared regulator tool}\label{modesmatch}
 

Let us look at the infrared behavior of the correlation functions associated to the field $\psi$. In four dimensions, and in the slow-roll approximation, the solution (\ref{psisoln}) is simplified by requiring that, for $\epsilon \rightarrow 0$, it matches the exact de Sitter solution. This implies that $E_{k}=1$ and $F_{k}=0$ and the integrand of the two-point function 
\bea
\langle\phi(x)\phi(x')\rangle=\int{d^{3}k\over (2\pi)^{3}}\,e^{i\vec k\cdot (\vec x-\vec x')}\psi(t,k)\psi^*(t',k)
\eea
at coincident points behaves  as $k^{-1-2\epsilon}$, as one can check by considering the small argument limit of the Hankel function \cite{abramo}
\bea\label{smallz}
H^{(1)}_{\mu}(z)\simeq {\Gamma(\mu)\over i\pi}\left(2\over z\right)^{\mu} \ .
\eea
Thus, the choice $E=1$, $F=0$ leads to an infrared divergence for the tensor perturbations but also for the scalar ones (it is sufficient to replace $\epsilon$ with $3\epsilon-\eta$ and keep in account that they are both much smaller than one). The latter can be avoided by considering a pre-inflationary phase, where the infrared modes are not divergent. In fact, according to a theorem formulated by Ford and Parker in \cite{ford}, infrared divergences cannot form in a expanding Universe. So, if they absent at the very beginning they will never appear later. One way to see this is to consider the following piece-wise solution to Eq.\ (\ref{psisoln})
 \bea\label{match}
\left.\begin{array}{ll}\psi_\nu(t)=a^{-1}_{\nu}(t)u_{\nu}(z)\ , &\qquad t<t_i \ ,\\\\\psi_\mu(t)=a^{-1}_{\mu}(t)\left[E_ku_\mu(z)+F_ku_\mu^*(z)\right] \ , & \qquad t_i<t\ ,\end{array}\right.
\eea
with $\mu=3/2+\epsilon$ and $\nu\neq \mu$. The scale factor changes at $t=t_{i}$, but we assume that $a_{\nu}(t_{i})=a_{\mu}(t_{i})\equiv a_{i}$ and $\dot a_{\nu}(t_{i})=\dot a_{\mu}(t_{i})$. In this way, the coefficients $E_{k}$ and $F_{k}$ can be uniquely determined by solving  the linear system
\bea\label{syst}
\left\{\begin{array}{c}\psi_\nu(t_i)=\psi_\mu(t_i)  \\\\ \dot\psi_\nu(t_i)=\dot\psi_\mu(t_i)\end{array}\right. \ .
\eea
By using the identity $\dot z(t) = -k/a(t)$, which holds whenever $\epsilon$ is constant, we find
\bea\label{EaFa}
\left.\begin{array}{c} E_k=ia_{i}\left(u_\mu^*\dot u_\nu-\dot u_\mu^* u_\nu\right)_{t=t_i}\ , \\\\ F_k=ia_{i}\left(\dot u_\mu u_\nu-u_\mu\dot u_\nu\right)_{t=t_i} \ .\end{array}\right.
\eea
With the help of the wronskian identity ($\gamma=\mu,\nu$)
\bea
u_\gamma(t)\dot u_\gamma^*(t)-u_\gamma^*(t)\dot u_\gamma(t)={i\over a_{\gamma}(t)} \ ,
\eea
we easily check that $|E_k|^2-|F_k|^2=1$.

To verify whether the divergence is still present, we first note that in the small $z$ limit (hence small $k$ limit), and according to the approximation (\ref{smallz}), we have $u_{\mu}\simeq -u^{*}_{\mu}$ thus $|\psi_{\mu}|^2\propto |u_\mu|^2 |E_k-F_k|^2 \propto k^{-2\mu}|E_k-F_k|^2$. In turn, as $|E_k-F_k|^2\sim k^{2(\mu-\nu)}$ for small $z_{i}\equiv z(t=t_{i})$, one has
\bea
\langle\phi^{2}(x)\rangle\propto \int dk k^{2(1-\nu)}\ .
\eea
This shows that the low-$k$ behavior of the integral is independent of $\mu$, namely of whether the Universe accelerates or decelerates after the transition at $t=t_i$.   It is also clear that the integral converges provided $\nu<3/2$, which excludes configurations where the Universe accelerates for $t<t_i$, such as de Sitter or slow-roll. On the contrary, if the Universe begins in a radiation-dominated phase, IR convergence is guaranteed. 

The question now is whether non-trivial values for $E_{k}$ and $F_{k}$ can lead to observational signatures in the power spectrum or in related quantities, such as the spectral index. It usually claimed that the effect of the matching is negligible as $F_{k}$ is very small for the scales of interest.  We remark that, generally speaking, this is not always a good motivation. In fact, even though the deviation form the standard, almost scale-invariant spectrum is small, the spectral index  can change significantly. Suppose in fact that the modified spectrum $P_{\rm mod}(k)$ is given in terms of the standard one $P_{\rm stand}\sim k^{-2\epsilon}$, $0<\epsilon\ll 1$ as
\bea
P_{\rm mod}(k)=P_{\rm stand}\left[1+f(k)\right]\ ,
\eea
for some function $f$ such that $f(k)\ll 1$ at the scales $k$ of interest. Current observations set the value of the scalar power spectrum to $(2.43\pm0.11)\times 10^{9}$ thus, any deviations below about $4\%$ would be invisible \cite{wmap5}. However, the spectral index is given by
\bea
n\equiv {d\ln P_{\rm mod}(k)\over d\ln k}=-2\epsilon+{k\over 1+f(k)}{ df(k)\over dk}\ ,
\eea
so it can be very sensible to small deviations if $k\,df/dk$ is large for the $k$ of interest. 

We now show that this is not the case for the situation at hand. In the case studied above, one should evaluate $|\psi_{\mu}|^2$ around the horizon exit. However, the coefficients $E_k$ and $F_k$ must be calculated at the time of the matching $z_{i}=z(t_{i})$. To achieve this, we  recall that $z_i=k(1+\epsilon_\mu)/a_iH_i$. Since during slow-roll $(1-\epsilon_\mu)\tau=-1/aH$, we find that $z_i=-k\tau_i$, where $\tau_i$ is the conformal time at the matching and the equality holds up to order $\epsilon^2$. On the other hand, the scales of interest crosses the horizon at a conformal time $\tau_k$ such that $k\tau_k=-(1+\epsilon)$. If we call $N$ the number of e-folds occurred between the time of the matching and the time of the horizon exit of the scales of interest, we find that $z_{i}=\tau_{i}/\tau_{k}=e^N$. Thus, at just few e-folds before the horizon exit, $z_{i}$ is a very large number. In this regime, with the help of Eq.\ (\ref{largearg}) we find that
\bea
E_{k}-F_{k}\simeq\exp\left[{i\pi\over 2}(\nu-\mu)\right] \ ,
\eea
so that $|E_{k}-F_{k}|^{2}$ is \emph{exactly} equal to one. This holds for both the tensor and the scalar power spectra, so the spectral indices are unaffected. The conclusion is that the pre-inflationary phase considered above cannot be detected with the present available data. We remark that the evaluation of the coefficients $E_{k}$ and $F_{k}$ for vanishing $k$ (and so vanishing $z_{i}$) is considered to study the IR behavior only. In this case, a non-vanishing $F_{k}$ guarantees a IR finite result. Instead, to compute the effects on the visible spectrum, one needs to evaluate these functions for large arguments. However, in this regime the weight of $F_{k}$ drops to zero.


\section{Dimensional transition and mode matching}\label{Dtrans}


The discussion in the previous section seems to definitively exclude the possibility of investigating a pre-inflationary phase modeled via mode matching. However, this conclusion depends crucially on the fact that $|E_{k}-F_{k}|^2=1$ at horizon exit, a direct consequence of keeping the same number of dimensions of space-time before and after the match. As we discussed in the introduction, we want to investigate what happens when the initial modes emerge from an effective two-dimensional space-time. During the pre-inflationary and  two-dimensional phase, we do not know the details of the theory, therefore we just assume that there exist one scalar degree of freedom, coupled to gravity through the parameter $\xi_{\nu}$, which propagates in an effective two-dimensional time-dependent metric. At the transition time $t_{i}$, the fluctuations of this scalar field set the initial conditions for both scalar and tensor perturbations of the four dimensional inflationary Universe, i.e. they fix the coefficients $E_{k}$ and $F_{k}$. 

Let us illustrate this mechanism with an explicit example. Consider the metric
\bea
ds^{2}=-dt^{2}+a^{2}dz^{2}+b^{2}(dx^{2}+dy^2),
\eea
where $a$ and $b$ are functions of time only. The Klein-Gordon equation (\ref{KG}) reads
\bea\label{anis}
\ddot \psi+(2H_b+H_a)\dot \psi+\left({k_\perp^2\over b^2}+{k_z^2\over a^2}+\xi R\right)\psi=0,
\eea
where
\bea
R&=&2\left(2\dot H_b+\dot H_a +3H_b^2+H_a^2+2H_aH_b  \right), \\\non
H_a&=&{\dot a\over a}, \quad H_b={\dot b\over b},\quad k_\perp^2=k_x^2+k_y^2
\eea
and where $\psi$ is the time-dependent part of the field
\bea
\phi(x)=\psi(t)\exp (ik_xx+ik_yy+ik_zz).
\eea
If we assume that
\bea
b(t)=\left\{\begin{array}{cc}{\rm const}=b_{0} & t<t_i \\a(t) & t>t_i\end{array}\right.
\eea
we find that, for $t<t_i$,
\bea\label{asimmodes}
\left[z^2{d^2\over dz^2}+z^2\left(1+{a^2k_\perp^2\over b^2_{0}k_z^2}\right)+{2\xi_<\over 1-\epsilon_a}\right]\psi(t)=0
\eea
where $z=k_z[\dot a(1-\epsilon_a)]^{-1}$, $\epsilon_a=-\dot H_a/H_a^2$, and $\xi_<$ denotes the value of the  coupling to gravity for  $t<t_i$. This equation is the same as Eq.\ (\ref{modeseq}) in $D=2$, provided $a^2k_\perp^2\ll b_{0}^{2}k_z^2$ for all $t<t_i$. Instead, for $t>t_i$, Eq.\ (\ref{anis}) coincide with  Eq.\ (\ref{modeseq}) in $D=4$. Another possibility is that the term
\bea
{z^2a^2k_\perp^2\over b_{0}^2k_z^2}={k_\perp^2\over b_{0}^2H_a^2(1-\epsilon^2_{a})^2}\simeq {\rm const}
\eea 
as in the case when the two-dimensional phase is of de Sitter type, so it can be absorbed into a redefinition of $\xi_<$, without imposing any constraint on the magnitude of $k_\perp$. Moreover, we will shortly see that the observable spectra are independent of the value of $\xi_<$, therefore the contribution of transverse modes is completely hidden in this case. Then, Eq.\ (\ref{asimmodes}) is again equivalent to Eq.\ (\ref{modeseq}) in $D=2$. In either case, we can match the solutions of the Klein-Gordon equation at $t=t_i$, by modeling the dimensional transition as a sudden decompactification of two transverse spatial directions. For $t<t_i$ the metric is effectively two-dimensional, but extra-dimensions are present and allow for tensor perturbations.

In the following analysis, we further assume that, after the matching, the Universe expands according to a slow-roll evolution and that perturbations are minimally coupled to gravity. We remind that the crucial assumption here is that the transition happens in such a way that there are no remnant preferred directions. As mentioned in the introduction, this can be achieved in few ways, and we assume that one of this mechanism is at work. According to the formalism presented above, we need to study the equations
\bea
\left.\begin{array}{ll}\psi_\nu(t)=u_{\nu}(z)\ , &\qquad t<t_i \ ,\\\\\psi_\mu(t)=a^{-1}_{\mu}(t)\left[E_ku_\mu(z)+F_ku_\mu^*(z)\right] \ , & \qquad t_i<t\ ,\end{array}\right.
\eea
where now
\bea
\nu=\sqrt{{1\over 4}-{2\xi_\nu\over 1-\epsilon_{\nu}}}\ , \qquad \mu\simeq {3\over 2}+\epsilon_{\mu}\ .
\eea
To find the coefficients $E_{k}$ and $F_{k}$, we need to solve the same system as in Eq.\ (\ref{syst}), with again the conditions that $a_{\nu}(t_{i})=a_{\mu}(t_{i})\equiv a_{i}$ and $\dot a_{\nu}(t_{i})=\dot a_{\mu}(t_{i})$. The results are
\bea
E_k&=&ia_i\Big[\left(\dot a u_\nu+a\dot u_\nu\right)u_\mu^*-au_\nu\dot u_\mu^*\Big]_{t=t_i,\,z=z_i}\ , \\\non
F_k&=&ia_i\Big[\,au_\nu\dot u_\mu-\left(\dot a u_\nu+a\dot u_\nu\right)u_\mu\Big]_{t=t_i,\, z=z_i}\ ,
\eea
and one can check  that $|E_k|^2-|F_k|^2=1$. By comparing this result with Eqs.\ (\ref{EaFa}), we see that it also depends on $\dot a(t_{i})$.

We are now ready to calculate the power spectra and the related spectral indices to estimate how much they deviate from standard results. In analogy with the previous calculations, we evaluate the coefficients $E_{k}$ and $F_{k}$ by using the approximation (\ref{largearg}), and the results are
\bea\non
E_k&\simeq&{i\over 2k}\left( \dot a_i-2ik \right)\, \exp\left[{i\pi\over 2}(\mu-\nu)\right]\ , \\
F_k&\simeq&-{\dot a_i\over 2k}\, \exp\left[-{i\pi\over 2}(\mu+\nu)+2iz_i\right]\ ,
\eea
which still satisfy $|E_k|^2-|F_k|^2=1$. As before, we have $|\psi_\mu|^2\simeq  a^{-2}|u_\mu(z_f)|^2|E_k-F_k|^2$ and, in the slow-roll approximation
\bea
|\psi_\mu|^2={GH^2\over \pi^2k^3}\left[1-{(1+\epsilon_\mu)\over z_i}\sin\left(2z_i-\epsilon_\mu \pi\right)+{\cal O}(z_{i}^{-2})\right]\ ,
\eea
where we used the standard mode normalization of \cite{lyth}.  As a result, the power spectrum  reads
\bea
P_{T}=64\pi G\left(H\over 2\pi\right)^{2}\left[1-{(1+\epsilon_\mu)\sin\left(2z_i-\pi\epsilon_{\mu}\right)\over z_i}+{\cal O}(z_{i}^{-2})\right]\ ,
\eea
and, anticipated above, the result does not depend on $\nu$, so it does not depend on the value of  $\xi_<$ of Eq.\ (\ref{asimmodes}) either.

We now consider the scalar perturbations and the associated spectrum. As we explained in the previous section, the scalar perturbation equation can be obtained from the tensor one by means of the identification (\ref{iden}).  Thus, the computation of the power spectrum is essentially the same as above, and we report only the result, namely
\bea\non
P_S={4\pi G\over\epsilon}\left(H\over 2\pi\right)^2\left[1-{(1+\epsilon_\mu)\sin\Big(2z_i-(3\epsilon_\mu -\eta)\pi\Big)\over z_i}+{\cal O}(z_{i}^{-2})\right]\ .\\
\eea

With these findings, we are now able to estimate the measurable quantities, in particular  the scalar-to-tensor ratio $r=P_T/P_S$ and the spectral indices. By expanding the ratio at large values of $z_i$, we find that
\bea
r=16\epsilon_\mu \left[1+{\pi(\eta-2\epsilon_{\mu})\cos(2z_i)\over z_i}\right]\ ,
\eea
which shows a damped oscillating correction to the usual result. As we explained before, we can characterize $z_{i}$ through the number of e-folds occurred between the matching time and the exit of interesting scales, namely $z_{i}=e^{N}$. It follows with already few e-folds,  the ratio $\cos(2e^{N})/e^{N}$ becomes very small: for $N>5$ this number drops below $10^{-2}$. Even in the extreme case when $z_{i}\simeq 1$, so that the transition happens around the horizon exit time, the correction would be of the order ${\cal O}(\epsilon^2)$.

Things can be different however concerning the spectral indices. In fact, for the tensor power spectrum  the index is given by
\bea
n_{T}\equiv{d\ln P_{T}\over d\ln k}=z{d\ln P_{T}\over dz}\simeq -2\epsilon_{\mu}-2(1+\epsilon_{\mu})\cos(2z_{i})\ ,
\eea
where the large $z_{i}$ limit has been taken. Similarly, for the scalar power spectrum we find
\bea
n_{S}-1\equiv{d\ln P_{S}\over d\ln k}\simeq2\eta-6\epsilon_{\mu}-2(1+\epsilon_{\mu})\cos(2z_{i})\ .
\eea
The other important parameter is the running of the spectral index, which, in the large $z_{i}$ limit, reads
\bea
n_{S}'\equiv {dn_{S}\over d \ln k }\simeq 4(1+\epsilon_{\mu})z_{i}\sin(2z_{i})\ .
\eea
At first sight, these results seem in complete disagreement with the data. The scalar spectral index oscillate rapidly in the range $-(2+8\epsilon-2\eta)<n_{S}<2+2\eta-4\epsilon$, while $n_{S}'$ oscillates and grows boundlessly with the number of e-folds. This interpretation is however wrong. In fact, the above results do not keep in account that we are not able to measure the index and the running with infinite precision. What we actually do is a sampling of these parameters, by means of intervals of width $|k_{2}-k_{1}|$. If the latter is larger than the wavelength of the oscillation, we will not see any difference in the spectral index with respect to the standard, constant result.
The outcome of a measurement of $n_{S}$ is in fact the integral average
\bea
\langle n_{S}-1\rangle&=&{1\over (k_{2}-k_{1})}\int_{k_{1}}^{k_{2}}dk\, n_{S}(z_{i})=\\\non
&=&2\eta-6\epsilon_{\mu}-2(1+\epsilon_{\mu}){\cos[\tau_{i}(k_{2}+k_{1})]\sin[\tau_{i}(k_{2}-k_{1})] \over \tau_{i}(k_{2}-k_{1})}\ .
\eea
A ``perfect'' sampling correspond to $k_{2}=k_{1}$ in which case the above equation yields again $n_{S}$. In our case, we have that $\tau_{i}=-z_{i}/k$ with $k$ of the order of magnitude of $k_{1}$ and $k_{2}$. A $1\%$ accuracy for $n_{S}$ corresponds to $(k_{2}-k_{1})/k$ of the order of $10^{-5}$ with 5 e-folds but of $10^{-24}$ with $50$ e-folds. Thus, unless the matching time is close to horizon exit, we will not see any effect in future data. The same conclusion holds for the measured running of the spectral index $\langle n_{S}'\rangle$. As the index itself appears to be constant, there will be no actual running. Again, one could consider the case when $z_{i}\simeq 1$, so that the corrections are not too large. However, we think that this situation is unphysical under another aspect. If the two-dimensional physics were still important at the time of the horizon exit, there would be large anisotropies frozen in the modes about to exit the horizon. In other words, the expansion would not have had the time to wash them out according to the mechanism proposed by Wald in \cite{wald}, and that would be clearly visible e.g. in CMB fluctuations. Moreover, we also feel that the correspondence between the time of the transition and the time of exit of the fluctuations that interest us would be quite an unjustified fine-tuning.


\section{Discussion}\label{conc}


In the previous sections we analyzed the impact of the matching on observable quantities, such as spectral index and running. The matching physically corresponds to a situation where effectively two-dimensional perturbations cross  into a four-dimensional space-time at a certain time $t_{i}$, which enter as a parameter in the theory. In fact, it explicitly appears in the expression of measurable quantities. However, we showed that the net effect of this new parameter is to add a very rapidly oscillating  component to the standard expression of the scalar index. We argued that this oscillation is not observable, unless we are in the not so realistic case where the dimensional transition occurs less than 5 e-folds before the horizon exit of the cosmological scales of interest. 

We presented an explicit model of dimensional transition, where space-time is always four-dimensional but two spatial directions are ``frozen'' before the matching time. In this case, tensor modes exist at all times and the mode matching is well-defined for both scalar and tensor perturbations. In particular, if the pre-matching phase is governed by an approximately de Sitter phase, we also showed that the transverse momenta associated to the  compact dimensions can be absorbed by a redefinition of the coupling to gravity and leave no trace in the final spectra.

It is interesting to note that our scenario can be extended to other kinds of modification of short distance gravity, which do not involve changes in the number of dimensions. One example is given by modified dispersion relations (MDR), initially inspired by analogue models of gravity \cite{unruh} and then considered on various context, from Unruh effect, to renormalization and trans-planckian cosmology, see e.g. \cite{MDR}. A simple way to introduce MDR consists in adding higher powers of $k^{2}$ in the ``mass'' term of Eq.\ (\ref{fmodeeq}). These are the results of adding higher order derivative operators in the Klein-Gordon equation for the test field (\ref{KG}). In terms of the variable $z$ this essentially amounts to write Eq.\ (\ref{modeseq}) as
\bea
 \left(z^2{d^2\over dz^2}+z^2+\gamma z^{2m}+{1\over 4}-\mu^2  \right)\left(a^{{D\over 2}-1}\psi\right)=0\ ,
 \eea
 where $\gamma>0$ sets a threshold for dispersion effects and $m>1$. We have seen before that the relevant regime where the matching occurs is for large $z$. Thus, the above equation simplifies to 
\bea
 \left(z^2{d^2\over dz^2}+\gamma z^{2m}\right)\left(a^{{D\over 2}-1}\psi\right)\simeq0\ .
 \eea
 In $D=4$, the solution reads
 \bea\non
 \psi(z)={\sqrt{z}\over a(t)}\, H^{(1)}_{1/2m}\left(\sqrt{\gamma}\,z^{m}\over m\right)\simeq {z^{(1-m)/2}\over a(t)}\exp\left({\sqrt{\gamma}\,z^{m}\over m} -{\pi\over 4m}-{\pi\over 4}\right).\\
 \eea
 We now recall that $z=k/(1-\epsilon)aH$, where $\epsilon$ is constant. If we also assume that, before the matching, $H$ is nearly constant, we see that the above expression for $\psi$ has a factor $a^{m/2-3/2}$ which reduces to the usual one when $m=1$. In fact, $m=1$ coincides with the relativistic linear dispersion case, at least in the large $z$ limit. If we now consider the system (\ref{syst}), we will have different powers of $a$ in the first and in the second equations. Therefore, we end up with a very similar situation as for the transition between two and four dimensions. Hence, our results can be extended also to the case when dimensionality does not change, but the dispersion is no longer linear before the matching, and we find that in this case there are no observable signatures. This is partially in contrast with the findings of \cite{brand}, where similar calculations were done. The oscillatory character of the spectrum was found also in this work, and is was interpreted as an imprint of transplanckian effects. However, we believe that the authors of  \cite{brand} failed to recognize that such oscillations are not observable with current techniques.

\ack

Many thanks to R. Durrer, M. Kunz, R.\ K.\ Jain, and G. Marozzi for valuable discussions.

\end{document}